# THE P- AND R-PROCESSES: REVIEWS AND OTHER VIEWS

M. ARNOULD, S. GORIELY AND M. RAYET

*IAA - ULB - Campus de la Plaine CP226, B-1050 Brussels, Belgium*

A review is presented of the p-process in Type II supernovae, one of its goals being to enlighten the changes in views on this nucleosynthesis mechanism since the work of Jean and Jim on the subject in 1975. Specific discussions are also devoted to cases of particular interest, like the light Mo and Ru stable isotopes, the rare nuclide $^{138}$La or the radionuclide $^{146}$Sm. Some comments of diverse natures are also made on the r-process. These considerations do not aim at really providing an exhaustive review of the many nuclear physics and astrophysics intricacies of this process. In contrast, they are hoped to complement or to put in perspective other views that are often expressed in relation with this nucleosynthesis mechanism

## 1  AN INTRODUCTION TO THE P-PROCESS IN TYPE II SUPERNOVAE

The p-nuclides are those neutron-deficient stable isotopes of the elements heavier than iron that cannot be produced by the s- or r-processes of neutron captures. In view of the fact that the number of research papers devoted to the p-process called for their production remains inferior to the number of p-nuclides after more than about 50 years of nuclear astrophysics research, we like to refer to the 'nuclear astrophysics p-nuts' when talking about the p-nuclides.[a] Either independently or in collaboration, Jean and Jim have had the good taste to contribute to the increase in the number of papers devoted to the p-process. In ref.[1], Jean put forth the possible spallogenic origin of selected p-nuclides, while in ref.[2], Jean and Jim pushed to its limits the original proposal[3] to produce the p-nuclides at temperatures in excess of $10^9$ K in the H-rich envelopes of Type II supernovae (SNII). Already at the end of the sixties, it was realized that such high temperatures were unlikely to be reached in the envelopes of massive star explosions. This is why one of the authors of this review (M.A.) proposed in his 1971 thesis work (see also ref.[4]) to locate the p-process in the deep O-Ne rich layers of massive stars either in their pre-supernova or supernova phases.

---

[a] The fact that the p-nuclides are much less abundant (by factors of the order of 100 to 1000) than the corresponding more neutron-rich isotopes in the solar system, which is the only astrophysical location to-date where they are observed, is certainly not a viable explanation for the scarcity of the efforts devoted to the understanding of their origin. The validity of this statement appears in a crystal-clear way when referring to the light elements Li, Be and B, to which myriads of astrophysics papers have been devoted (a scan of the contributions to this conference is quite convincing in this respect)

The SNII p-nuclide synthesis has been the most actively studied since then, and is certainly to-date the most developed model and the most successful one in reproducing the p-nuclide content of the bulk solar system material (e.g. ref.[5], hereafter RAHPN, and references therein). The extensive RAHPN calculations rely on SNII explosion models for stars with initial masses in the $13 \lesssim M \lesssim 25$ M$_\odot$ range[6]. The p-nuclides are produced in layers with explosion temperatures peaking in the $(1.8-3.3) \times 10^9$ K range. The nuclear flow in these conditions is dominated by photodisintegrations of the $(\gamma,n)$, $(\gamma,p)$ or $(\gamma,\alpha)$ types, complemented mainly with some neutron captures. Figure 1 shows the p-process overproduction factors obtained for the 25 M$_\odot$ model star selected by RAHPN, and for two sets of reaction rates. It appears that about 60% of the overproductions displayed in Fig. 1 fit the solar system composition within a factor 3. This conclusion is in general not drastically affected either by the selected set of reaction rates, or by the stellar mass, at least in the range explored by RAHPN. However, the predictions are not free from shortcomings, some of them being discussed below.

## 2 THE PUZZLE OF THE LIGHT Mo AND Ru ISOTOPES

One of the most embarassing problems which is evidenced by Fig. 1 concerns the severe underproduction of the light Mo ($^{92-94}$Mo) and Ru ($^{96-98}$Ru) isotopes, this conclusion being robust to various changes in the input physics (see RAHPN). Exotic solutions have been proposed to remedy this puzzle, calling in particular for accreting neutron stars or black holes (e.g. ref. [9]). The level of the contribution of such sites to the solar system content of the p-nuclides is clearly impossible to assess in any reliable way. In contrast, we have emphasized many times in the past that the problem might just be due to some misrepresentation of the production in the He-burning cores of massive stars of the s-nuclides which are the seeds for the p-process. This 'non-exotic' idea has been put on quantitative grounds recently[10]. More specifically, the seed abundances from ref.[11] used to construct Fig. 1 are replaced by results obtained with the 'adopted' NACRE reaction rates, except for the rate $r_{22}$ of $^{22}$Ne$(\alpha,n)^{25}$Mg which is the main neutron source in the considered stars. This rate is in fact varied in the range of uncertainties defined in the NACRE compilation.

Use of the NACRE 'adopted' value for $r_{22}$ leads to the classical 'weak' s-process component pattern exhibiting a decrease of the overproduction (with respect to solar) of the s-nuclides by a factor ranging from about 100 to unity when the mass number $A$ increases from about 70 to 100. This 'canonical' picture changes gradually with an increase in the neutron production rate, the overproduction of heavier and heavier s-nuclides increasing steadily. More specifically, adoption of the NACRE upper limit for $r_{22}$ leads to s-nuclide overproduction factors from about $10^3$ to $10^4$ in the $70 \lesssim A \lesssim 90$ mass range, decreasing to values around unity for mass numbers as high as about 150 (ref.[10]). These results obtained with increased $r_{22}$ values may shake some aspects of a long tradition in our views concerning the s-process. In particular, massive stars appear to be able to produce quite substantial amounts of s-nuclides around Ba. At the same time, Asymptotic Giant Branch stars synthesize s-elements in the Zr region. It is our opinion that none of these predictions can really act as a deterrent to $r_{22}$ substantially in excess of the NACRE adopted value. Quite on the contrary, some specific problems might even find a solution, like the (admittedly quite uncertain) Ba abundance in SN1987A[10].

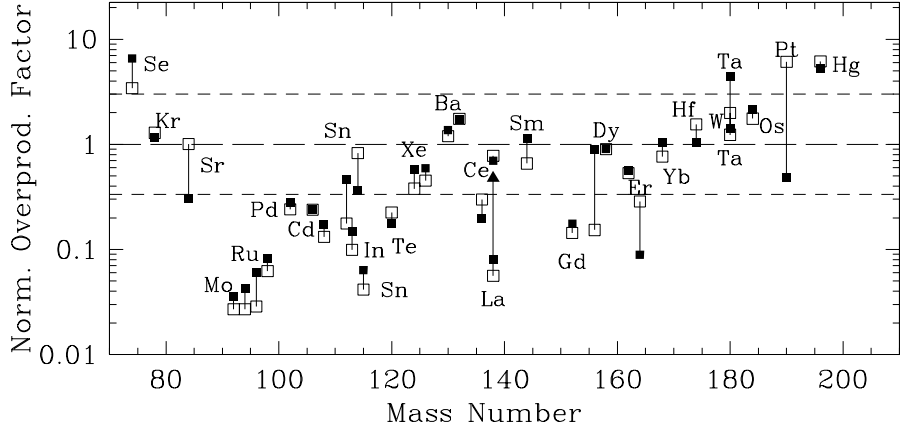

Figure 1. Normalized p-nuclide overproduction factors (with respect to solar) in the 25 $M_\odot$ star model adopted by RAHPN. The open squares are obtained with the reaction rates considered by RAHPN, while the black squares relate to the use of the 'adopted' rates from the NACRE compilation[7] and of the Hauser-Feshbach predictions from the code MOST[8]. In both cases the s-process seed abundances are those of ref.[11]. The NACRE and MOST rates are available in the Brussels Nuclear Astrophysics Library (http://www-astro.ulb.ac.be). The black triangle at $^{138}$La indicates the yield increase by a factor of 6 obtained by adopting an evaluated upper (lower) limit of the rate of $^{138}$La production (destruction) (see Sect. 3)

Use of the s-process seed abundances obtained with the NACRE adopted and upper values for $r_{22}$ leads to the 25 $M_\odot$ p-nuclide overproduction factors

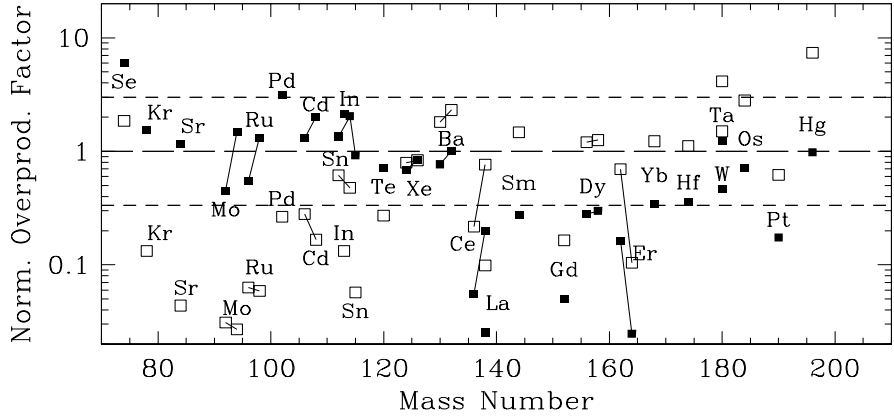

Figure 2. Same as Fig. 1 (black squares), but for s-process seeds obtained with the NACRE rates. Open and black squares correspond to the NACRE 'adopted' and 'upper' $^{22}$Ne$(\alpha,n)^{25}$Mg rates, respectively. Solid lines join different p-isotopes of the same element

shown in Fig. 2. Changes in the shape of the p-nuclide abundance distribution are clearly noticeable. In particular, the Kr-Sr-Mo-Ru trough observed for the NACRE adopted $r_{22}$ disappears when the upper limit for that rate is considered. Details about this effect as well as other interesting by-products of the increase of $r_{22}$ are discussed in ref.[10].

## 3  SOME OTHER SPECIFIC QUESTIONS OF p-PROCESS NATURE

Other important questions directly connected with the p-process have been the subject of specific research. One of them concerns the experimental study of the $^{144}$Sm$(\alpha,\gamma)^{148}$Gd rate and the analysis of its implication in the development of a possible nucleo-cosmochronology of the p-process. The measurements have proved to be of substantial nuclear physics interest, and have been used to estimate the p-process production ratio of $^{144}$Sm to the radionuclide $^{146}$Sm ($t_{1/2} \approx 10^8$ y). From the confrontation of these evaluations with the corresponding solar system abundance ratio inferred from the analysis of primitive meteorites, one might hope to estimate the age of the p-nuclides. Large uncertainties still remain in this chronology in spite of the nuclear physics advances reported above. The interested reader can find a detailed account of this study in ref.[12].

On the other hand, some study has been devoted to $^{138}$La which is, along with $^{180}$Ta, an odd-odd p-nuclide with very low solar system abundances. In spite of their scarcity, the very origin of these two species has long been a nucleosynthesis puzzle. As discussed by RAHPN and as shown in Figs. 1 and 2, $^{180}$Ta is found to be a natural p-process product. In contrast, $^{138}$La is systematically underproduced in all p-process calculations performed so far. This situation has triggered an examination of the sensitivity to nuclear uncertainties of its predicted yields, which result from a subtle balance between its production by $^{139}$La$(\gamma, n)^{138}$La and its photodisintegration at temperatures around $T \simeq 2.4 \pm 0.1\ 10^9$K. No experimental cross sections are available, so that the $^{138}$La yields of Fig. 1 entirely rely on the predictions of the code MOST. Related reasonable uncertainties in the neutron capture rates on $^{138}$La and on $^{137}$La (from which the $^{138}$La destruction rate is evaluated) at typical p-process temperatures are found to amount to factors of about 10 and 3, respectively. If the corresponding lower limit of the $^{138}$La destruction rate and upper limit of its production rate are adopted, the $^{138}$La yield from the 25 M$_\odot$ supernova shown if Fig. 1 is increased by a factor of about 6. This puts the $^{138}$La overproduction at the same level as the one of the neighboring p-nuclides. Nuclear physics measurements are needed in order to shed light on a possible nuclear solution to the $^{138}$La mystery.

Other studies relating to the p-process are currently pursued. They concern in particular the computation of this synthesis mechanism in C-deflagrating white dwarfs (in collaboration with J. Jose, Barcelona), or in the framework of 2-D simulations of the O-burning zone of a 20 M$_\odot$ presupernova star (in collaboration with D. Arnett and co-workers, Tucson).

## 4 SOME VIEWS ABOUT THE r-PROCESS AND RELATED QUESTIONS

Since the seminal work of ref.[13], as well as of some early research by Jim and his collaborators, much nuclear physics and astrophysics efforts have been devoted to the r-process aimed at accounting for the production of the neutron-rich stable isotopes of the elements heavier than iron. In spite of that, one is clearly left with a variety of difficult questions concerning the very site(s) of the r-process, and the properties of the very exotic neutron-rich nuclides it involves. In the following, we raise some questions and make some short sketchy comments, a fraction of which might be considered by certain researchers in the field as highly unorthodox:

**Question 1.** *What is (are) the true astrophysical site(s) of the r-process?*
**Comment 1.** The lack of answer to this essential question prevents any firm (sometimes far-reaching) conclusion to be drawn in relation with the r-process.

**Question 2.** *What are the true properties of the highly exotic neutron-rich nuclides involved in the r-process?*
**Comment 2.** One cannot derive the properties of exotic nuclei and identify the 'best' nuclear models from a confrontation between r-nuclide observations and predictions from a (toy) r-process model. This exercise (which is performed by some) is a kind of highly complex inverse problem the solution of which is unreliable, and even possibly non-unique.

**Question 3.** *Can one build reliable r-process cosmic clocks (which go beyond age limits, or are more reliable than other dating methods) from long-lived actinides observed in the solar system?*
**Comment 3.** In several papers, Jim and his collaborators have attempted to answer this important question. Our views may be summarized as follows (e.g. ref.[14] for references): the classical r-process chronometers ($^{232}$Th-$^{238}$U, $^{235}$U-$^{238}$U) are most likely quite poorly reliable if all possible sources of uncertainties (arising from the r-process production ratios, the models for the chemical evolution of the Galaxy, ...) are taken into account in a fair way. In fact, the best hope in the field may well come from the $^{187}$Re-$^{187}$Os pair, even if it is not free from uncertainties. This is due to the s-process nature of $^{187}$Os (apart from the radiogenic contribution from $^{187}$Re). There is also reasonable hope for the nuclear problems raised by that pair to find answers in the nuclear physics laboratory in a foreseeable future.

**Question 4.** *Can one build a reliable cosmic clock from Th and other r-nuclides observed in some very metal-poor stars?*
**Comment 4.** The recent observation of r-nuclides, including Th, in some ultra-metal-poor halo stars (see the review by C. Sneden in these Proceedings) has been the subject of some excitement and of several studies by Jim and some of his collaborators. Quite generally, these Th observations are considered to provide a r-process clock of unmatched quality. This conclusion relies heavily on some critical assumptions. One of them concerns the 'uniqueness' of the r-process in view of the striking similarity of the r-nuclide abundances in the analyzed stars and the corresponding solar system abundances. This view is challenged in ref.[15]; see also S. Goriely, these proceedings). On the other hand, and even if the r-process uniqueness is taken for granted, it is close to impossible to evaluate the level of Th production in the r-process with the required accuracy (i.e. within about 16%; ref.[16] and S. Goriely, these pro-ceedings). The uncertainties in the predictions of these yields related to the difficulties in the involved nuclear physics and their impact on age determinations are examined in detail in ref.[16] (see also S. Goriely, these Proceedings). Other possible sources of embarassment for a Th-based chronology are discussed and analyzed in the same references. A note of hope might come from a reliable determination of the Th/U ratio at the surface of ultra-metal-poor stars.

## 5   A FINAL COMMENT

There is no doubt that Jean and Jim have contributed in an important way to the understanding of the p- and r-processes. They have quite fortunately left unanswered questions, so that there has been room for others to tackle some interesting problems. More generally, they have been actively involved in the shaping of many chapters of nuclear astrophysics in a very professional manner, but also with a precious sense of humanism.